\documentclass[aoas]{imsart}

\RequirePackage{amsthm,amsmath,amsfonts,amssymb}
\RequirePackage[authoryear]{natbib}
\RequirePackage[colorlinks,citecolor=blue,urlcolor=blue]{hyperref}
\RequirePackage{graphicx}

\usepackage{enumerate}
\usepackage{url} 

\usepackage{physics}
\usepackage{svg}
\usepackage{mathrsfs}
\usepackage{xcolor}

\usepackage[algoruled,vlined]{algorithm2e}

\startlocaldefs
\theoremstyle{plain}

\theoremstyle{remark}

\newcommand{\DDg}[4]{{#1}_{#2}^{#3}(#4)} 
\newcommand{\DD}[3]{{#1}_{#2}^{#3}} 

\endlocaldefs

\begin{document}

\begin{frontmatter}
\title{Investigating the HIV Epidemic in Miami Using a Novel Approach for Bayesian Inference on Partially Observed Networks}
\runtitle{Bayesian Inference on Partially Observed Networks}

\begin{aug}
\author[A]{\fnms{Ravi}~\snm{Goyal}\ead[label=e1]{r1goyal@health.ucsd.edu}\orcid{0000-0002-0358-2435}},
\author[B]{\fnms{Kevin}~\snm{Nguyen}\ead[label=e2]{kevin@ohsu.edu}},
\author[C]{\fnms{Victor}~\snm{De Gruttola}\ead[label=e3]{degrut@hsph.harvard.edu}},
\author[A]{\fnms{Susan J}~\snm{Little}\ead[label=e4]{slittle@health.ucsd.edu}},
\author[D]{\fnms{Colby}~\snm{Cohen}\ead[label=e5]{colby.cohen@flhealth.gov}}
\and
\author[A]{\fnms{Natasha K}~\snm{Martin}\ead[label=e6]{natasha-martin@health.ucsd.edu}}
\address[A]{Division of Infectious Diseases and Global Public Health,
University of California San Diego\printead[presep={,\ }]{e1,e4,e6}}

\address[B]{Department of Medical Informatics and Clinical Epidemiology,
Oregon Health \& Science University\printead[presep={,\ }]{e2}}

\address[C]{Division of Biostatistics,
University of California San Diego\printead[presep={,\ }]{e3}}

\address[D]{Bureau of Communicable Diseases,
Florida Department of Health\printead[presep={,\ }]{e5}}
\end{aug}

\begin{abstract}
Molecular HIV Surveillance (MHS) has been described as key to enabling rapid responses to HIV outbreaks. It operates by linking individuals with genetically similar viral sequences, which forms a network. A major limitation of MHS is that it depends on sequence collection, which very rarely covers the entire population of interest. Ignoring missing data by conducting complete case analysis--which assumes that the observed network is complete--has been shown to result in significantly biased estimates of network properties. We use MHS to investigate disease dynamics of the HIV epidemic in Miami-Dade County (MDC) among men who have sex with men (MSM)--only $30.1\%$ have a reported sequence. To do so, we present an approach for making Bayesian inferences on partially observed networks. Through a simulation study, we demonstrate a reduction in error of $43\%-63\%$ between our estimates and complete case analyses. We estimate increased mixing between MSM communities in MDC, defined by race and transmission risk compared to the results based on complete case analysis. Our approach makes use of a flexible network model--congruence class model--to overcome the high computational burden of previously reported Bayesian approaches to estimate network properties from partially observed networks.
\end{abstract}


\end{frontmatter}

\section{Introduction}

The ability to respond rapidly to infectious disease outbreaks is critical to ending current and future epidemics. This is particularly relevant to ongoing U.S. initiatives such as the Ending the HIV Epidemic (EHE) initiative\citep{fauci2019ending} and the Hepatitis C Virus Elimination Plan, both of which include targets to reduce incidence of these infections. Responses by public health departments will require leveraging a range of public health strategies that identify gaps in prevention and care services. Molecular HIV Surveillance (MHS) has be a key approach to identifying outbreaks;\citep{oster2021hiv} it operates by linking individuals with genetically similar viral sequences. Investigation of the collection of such links, which form networks that we refer to as viral genetic linkage (VGL) networks, can identify areas of rapid transmission as well as aid in overall understanding of disease dynamics within a population. This paper uses MHS to investigate HIV transmission dynamics for Miami-Dade County (MDC). Of particular interest is the nature of HIV spread within MDC across communities defined by race/ethnicity and transmission risk. However, analyzing MHS data from MDC and other settings presents challenges in that there exists a considerable amount of missing sequence information, which results in the VGL network being only partly observed. To address this issue, we develop an approach to conduct Bayesian inference on partially observed networks that overcomes limitations of currently available methods.

Due to the viral evolution, linking individuals based on a threshold of their viral genetic similarity provides information about transmission between pairs of individuals who either directly infected each other or did so indirectly through a small subset of intermediaries. MHS avoids relying on case interviews and self-reported surveys, which are both prone to generating biases in estimates.\citep{helleringer2011reliability} In addition, MHS may reveal connections of which even the individuals themselves may not be aware. MHS has provided unique insights into HIV outbreaks. For example, analyses of sequences in Cabell County, West Virginia in 2018–2019 allowed estimation of  the transmission rate and timing of infections.\citep{mcclung2021response} In San Antonio, Texas, MHS identified 27 persons with similar HIV molecular sequences, which resulted in identification and further investigation of the HIV outbreak.\citep{oster2018molecular} A key limitation of investigating disease dynamics using MHS is that precision and robustness of analyses depend depend on completeness of sequence collection and reporting.\citep{oster2021hiv}
 
Ignoring missing data by conducting complete case analysis has been shown to result in significant biases in the estimation of network properties\citep{kossinets2006effects, smith2013structural, smith2017network} and parameters for network models simulating disease epidemics.\citep{krause2020missing, smith2022network} Recently, model-based approaches have made use of exponential random graph models (ERGMs)\citep{robins2007introduction} to analyze incomplete network data; frameworks for such analyses exist for both maximum likelihood estimation\citep{handcock2010modeling} and Bayesian inference\citep{caimo2011bayesian}. Model-based approaches have been shown to produce more reliable estimates than simple non-model-based approaches, such as using available cases, null-tie, and reconstruction.\citep{krause2019missing} ERGMs have many strengths that make them a useful and popular class of network models, such as the ease of making inference that results from their being within the exponential family \citep{robins2007introduction,  lusher2013exponential}. Nonetheless, ERGMs suffer from several weaknesses that limit their usefulness for investigating incomplete network data. In particular, these weaknesses include high computation costs when conducting Bayesian inference\citep{caimo2011bayesian, koskinen2013bayesian}--the more natural paradigm for addressing missing data. Specifically, Bayesian inference of ERGMs in the presence of missing data suffer from a double intractability issue--the normalizing constant for the  likelihood and posterior are both infeasible to compute except for small networks.\citep{koskinen2013bayesian}

We introduce a method for performing Bayesian inference on partially observed network data, utilizing a versatile network model known as the congruence class model (CCM) for networks.\citep{goyal2014sampling} CCMs encompass a wide range of models, including popular network models like the Erd\H{o}s-R\'{e}nyi (ER) model, the stochastic block (SB) model, and several Exponential Random Graph Models (ERGMs), as special instances.\citep{goyal2023estimating} Our proposed approach of using CCMs to address missing data provides two important advantages. The first is greater flexibility in modeling the probability distribution for network properties in terms of the functional form and number of parameters.\citep{goyal2023framework} The second is that the use of CCMs permits us to overcome the double intractability of parameter estimation that hinders current methods from scaling to larger networks. These pair of advantages may seem competing--a more general network model and decreased computational burden; however, recent advances in addressing the graph enumeration problem makes this possible.\citep{goyal2022general} Therefore, in the proposed approach, using CCMs permits investigation of more complex network properties, such as the entire degree distribution, even for large network (tens of thousands of individuals), which was previously infeasible. 

The paper is organized as follows. The next section (Section~\ref{sec:background}) introduces terminology and notation used in the paper as well as details about CCMs and about relevant previous research on inference of missing network data. Section~\ref{sec:methods} presents our approach of using CCMs for estimating the network properties in the presence of missing data. Section~\ref{sec:simulation} presents an extensive simulation study that investigates the potential of our approach. Section~\ref{sec:application} investigates HIV dynamics using MHS for MDC. Specifically, we provide a background on the HIV epidemic in MDC and available data, such as the level of MHS sequence coverage by demographic characteristics. The section also includes our investigation of disease dynamics in MDC related to the nature of HIV spread across racial and transmission risk communities. The paper concludes with a discussion (Section \ref{sec:discussion}).  

\section{Background} \label{sec:background}

\subsection{Terminology} 

For consistency, the notation presented follows previous research.\citep{goyal2023estimating} Let $g_{c} = (V,E)$ be the entire network, where $V = \{v_1,\ldots,v_n\}$ is the set of individuals in the population of interest and $E = \{E_{v_{i},v_{j}}\}$ is the set of indicators designating the presence $(E_{v_{i},v_{j}} = 1)$ or absent $(E_{v_{i},v_{j}} = 0)$ of a link between $v_i,v_j \in V$. Let $g_{o}$ be a partially observed network of $g_c$; therefore $g_o \subset g_c$. We assume that all individuals are observed, but links between the individuals may not be observed. Let $I_{v_{i},v_{j}}$ be an indicator of whether the value $E_{v_{i},v_{j}}$ is known ($I_{v_{i},v_{j}} = 1$) or not ($I_{v_{i},v_{j}} = 0$). Let $g_{u}$ be the unobserved portion of $g_{c}$, i.e., $g_c = g_o \cup g_u$. For simplicity, we use the notation $g$ to denote an arbitrary network.  

The number of links between an individual and others is referred to as the degree of that individual. We denote the degree for individual $v_i$ as $\DDg{d}{v_i}{}{g}$. The frequency of the degrees can be summarized into a vector referred to as the degree distribution, denoted as $\DDg{D}{}{}{g}$. The $j^{th}$ entry of the vector represents the number of individuals having degree $j$, e.g., $\DDg{D}{j}{}{g} = \sum_{i=1}^{n} I_{\{\DDg{d}{v_i}{}{g} = j\}}$. In addition to degree, individuals have other characteristics. Based on these characteristics, we group individuals, denote the number of distinct classifications as $q$. For example, in our investigation of HIV disease dynamics in MDC, individual-level characteristics include four race/ethnicity populations as well as if they reported injection drug use; therefore, $q=8$. We represent the  classification for individual $v_i$ in network $g$ as $\DDg{m}{v_i}{}{g}$ represent a discrete. The frequency of the links between individuals based on their classification can be summarized into a $q \times q$ symmetric matrix, which we refer to as a classification mixing matrix and denote as $\DDg{MM}{}{}{g}$. The entry $\DDg{MM}{k,l}{}{g}$ is the total number of link between an individual with classification $k$ and an individual with classification $l$. 

\subsection{Congruence class model}

Although CCMs have been previously been described,\citep{goyal2014sampling, goyal2023estimating} for convenience of the reader, we provide details and an illustration below. A CCM establishes a probability mass function (PMF) over the space of all networks with $n$ individuals, denoted as $\mathscr{G}_n$. The PMF, denoted as $P_{\mathscr{G}_n}(g \vert \theta)$, determines the likelihood of network $g$ given a vector of model parameters $\theta$. 

To specify $P_{\mathscr{G}_n}(g \vert \theta)$, we partition $\mathscr{G}_n$ into congruence classes. These classes are delineated by an algebraic mapping, denoted as $\phi$, from $\mathscr{G}n$ to network summary statistics of interest (e.g., degree distribution or characteristic mixing matrix). Let $c_{\phi}(x) = \{g : \phi(g) = x, g \in \mathscr{G}_n\}$ represent the inverse image corresponding to $\phi$. These inverse images are termed congruence classes.\citep{goyal2014sampling} The probability distribution specified on these congruence classes induces the PMF, $P_{\mathscr{G}_n}(g \vert \theta)$, for CCMs. We denote the PMF on the congruence classes as $P_{\phi}(x \vert \theta)$, which equals the sum of probabilities for all networks $g \in c_{\phi}(x)$, i.e.,

\begin{equation} \label{eq:networkprob_static_pc}
P_{\phi}(x \vert \theta) = \sum_{g \in c_{\phi}(x)} P_{\mathscr{G}_n}(g \vert \theta).
\end{equation}

\noindent CCMs posit that all networks within a congruence class share identical probabilities, a premise also adopted by commonly utilized network models such as ER, SB, and ERGMs. For instance, when the network summary statistic under consideration is the number of edges, all networks featuring the same edge count will have equivalent probabilities. Therefore, the induced probability distribution on $\mathscr{G}_n$ for a CCM is the following:

\begin{equation} \label{eq:networkprob_static}
P_{\mathscr{G}_n}(g \vert \theta) =\left(\frac{1}{\vert c_{\phi}(\phi(g)) \vert} \right) P_{\phi}(\phi(g) \vert \theta),
\end{equation}

\noindent where $\vert c_{\phi}(x) \vert$ denotes the cardinality of the congruence class $c_{\phi}(x)$.

\subsubsection{Practical considerations}

For $P_{\phi}(x \vert \theta)$ to be a proper probability function, it is necessary that the sum over all congruence classes equals 1, i.e., 

\begin{equation}
\sum_{u \in U_{\phi}} P_{\phi}(u \vert \theta)  = 1,
\end{equation}

\noindent where $U_{\phi} = \{x \colon g \in \mathscr{G}_n, \phi(g)=x\}$. For many specifications of $\phi$, it is non-trivial to meet this condition of a proper probability function. Therefore, in general, we have:

\begin{equation} \label{eq:networkprob_static_pc_QW}
P_{\phi}(\phi(g) \vert \theta) = \frac{Q_{\phi}(\phi(g) \vert \theta)}{W_{\phi}(\theta)},
\end{equation}

\noindent where,

\begin{equation}
W_{\phi}(\theta) =\sum_{u \in U_{\phi}} Q_{\phi}(u \vert \theta)  
\end{equation}

\noindent is a normalizing factor for the unnormalized mass $Q_{\phi}(\theta)$, over the set of congruence classes $U_{\phi}$. 

There are settings in which we only require knowing $P_{\mathscr{G}_n}(g \vert \theta)$ up to a normalizing constant, such as when we are sampling networks based on Equation~\ref{eq:networkprob_static} for a given $\theta$.\citep{goyal2014sampling} In these settings, specifying an unnormalized distribution provides an adequate and convenient way to assign $P_{\phi}(x \vert \theta)$. Below we provide an illustration of this advantage; other examples that do as well have been described elsewhere.\citep{goyal2014sampling, goyal2023estimating} For other settings,  it is necessary to consider $W_{\phi}(\theta)$. In this paper, we consider both situations and discuss $W_{\phi}(\theta)$ where appropriate.

\subsubsection{Illustration of a CCM}

Below, we present an illustration of a CCM that provides details on its specification. In particular, we highlight the flexibility in setting the probability distribution on the congruence classes, $P_{\phi}(\phi(g) \vert \theta)$. In addition, the illustration demonstrates the accuracy of the methods underpinning CCMs for generating networks that are consistent with the investigator specified probability distribution on the congruence classes.

For the illustration, we generate networks using CCMs that consist of $100$ individuals. We investigate degree distribution as the network property of interest, i.e., $\phi(g) = \DDg{D}{}{}{g}$. Therefore, we set $P_{\phi}(\phi(g) \vert \theta)$--the probability distribution for congruence classes defined by distinct degree distributions. CCMs place minimal restrictions on the specification of $P_{\phi}(\phi(g) \vert \theta)$; for example, the distribution can be  Poisson, power-law, or non-parametric. For this illustration, we assume that the degrees  follows a multinomial distribution with parameter vector $\theta$, where $\theta_i$ represents the probability of a node having degree $i$. We specify $\theta$ using a negative binomial distribution fitted to the University of California San Diego Primary Infection Resource Consortium (PIRC)--an observational cohort of people living with HIV.\citep{le2013enhanced} Specifically, we set:

\begin{equation} \label{eq:CCM_example_theta}
\theta_i \sim \frac{NB(i, size = 1.02, \mu = 6.19)}{z}
\end{equation}

\noindent for $i \in {0,\ldots,99}$ with $z$ as a normalizing constant.\citep{goyal2023estimating} The latter is necessary to address the issue that the negative binomial is a distribution over all non-negative integers, including values above $99$ (maximum degree). Note that $z$ differs from $W_{\phi}(\theta)$ and is easy to calculate.

As mentioned in Section~\ref{sec:background}, to sample networks, the probability of a network only needs to be specified up to a normalizing constant when given $\theta$. Therefore, using a multinomial distribution on the congruence classes provides a convenient approach to specifying a probability mass function on the space of networks; this approach circumvents the need to delineate all degree distributions that are graphical, i.e., the set $U_{\phi}$.

\begin{figure}
\centering
\includegraphics[width=400pt]{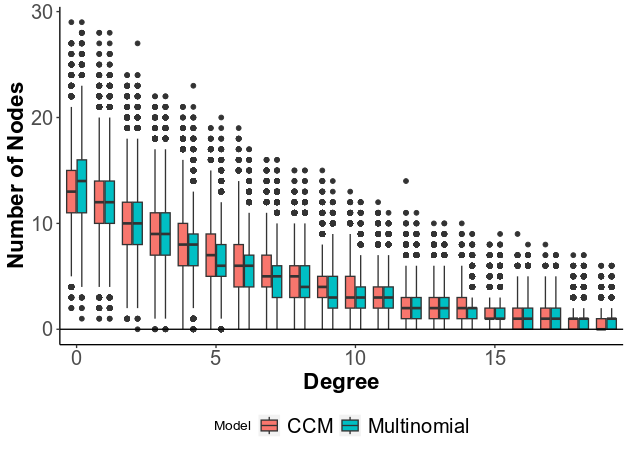}
\caption{Box plots comparing the number of individuals with degrees $0$ to $20$ for the networks generated using the CCM (red) and the samples from the multinomial (blue). Each box plot shows the number of individuals with a particular degree.}
\label{fig:CCM_illustration}
\end{figure}

For our simulation study, we generate a collection of $m = 50,000$ networks based on Equation~\ref{eq:networkprob_static} and the multinomial distribution described above. For each network $g_k$ where $k \in \{1,\ldots,50,000\}$, we summarize the number of individuals with each of the degrees from $0$ to $99$, i.e., calculate $\DDg{D}{i}{}{g_k}$. Similarly, we generate $m$ random samples of size $100$ from the multinomial distribution with a $\theta$ specified in Equation~\ref{eq:CCM_example_theta}. These $m$ samples represent our target distribution for the degree distribution, i.e., the distribution used to specify $P_{\phi}(\phi(g) \vert \theta)$. If the methods underpinning the network generation process for CCMs yield accurate results, we should see strong agreement between the degree distribution of networks generated by the CCM and the samples from the multinomial. Figure~\ref{fig:CCM_illustration} provides a comparison between the number of individuals with degrees $0$ to $20$ for the networks generated using the CCM (red boxplots) and the samples from the multinomial (blue boxplots); we observe few individuals with degree above 20. The distributions of degrees from the networks generated by the CCM and by the multinomial align closely, i.e. the medians and variances for each degree are very similar. For more complex illustrations, see \cite{goyal2014sampling}.

\subsection{Inference with Missing Network Data} \label{sec:ignorable}

It has been shown that inference for the parameter $\theta$ may be based on the face-value likelihood $\sum_{g_u} p(g_{o}, g_u|\theta)$ under the condition that the missing data mechanism is ignorable.\citep{handcock2010modeling, koskinen2013bayesian} Specifically, ignorability implies that (1) the conditional distribution of $I_{v_{i},v_{j}}$, the indicator of whether the value of link $E_{v_{i},v_{j}}$ is known, only depends on the observed part of data--which, in our case, includes both links and individual-level covariates; and (2) the CCM parameter $\theta$ and the parameter(s) for the observation process are distinct.\citep{koskinen2013bayesian} 

\section{Methods} \label{sec:methods}

We propose a Bayesian approach for estimation of the posterior distribution, denoted as $\pi(\theta \vert I, g_o)$, for parameters of a CCM, $\theta$, given the missing data indicator $I$ and partially observed network $g_o$. This posterior distribution can be augmented to include the unobserved portion of the network, $g_u$;\citep{koskinen2013bayesian} this augmentation results in the following posterior distribution:

\begin{equation} \label{eq:posterior_distr}
\pi(\theta, g_u \vert I, g_o) \propto P_{\mathscr{G}_n}(g_c = g_u \cup g_o \vert g_o, \theta) \pi(\theta).
\end{equation}

\noindent The probability of sampling an unobserved network $g_u$ given the observed part $g_o$, denoted as $P(g_u | g_o)$, is the same as the probability of sampling the complete network $g_c$ given $g_o$. That is $P(g_u | g_o) = P(g_c | g_o)$. This equivalence follows from the 1-1 mapping between $g_u$ or $g_c$ given $g_o$.

\subsection{Derive Posterior Distribution}

In order to generate samples from the posterior distribution described in Equation~\ref{eq:posterior_distr}, we employ a Gibbs sampler--a specific Markov Chain Monte Carlo (MCMC) method. In each iteration, the parameters $g_u$ and $\theta$ are updated sequentially, yielding a set of networks and network model parameter values that align with the partially observed network data.

\subsubsection{Update Unobserved Portion of Network: $g_u$}

Based on the posterior distribution shown in Equation~\ref{eq:posterior_distr}, the full conditional distribution for $g_u$ can be shown to be the following:\citep{koskinen2013bayesian} 

\begin{equation}  \label{posterior_dist_CCM_g}
    P(g_u \vert \theta, g_{o})) \propto \left(\frac{1}{\vert c_{\phi}(\phi(g_u \cup g_o)) \vert} \right) P_{\phi}(\phi(g_u \cup g_o) \vert g_o, \theta).
\end{equation}

\noindent Since $g_u \cup g_o = g_c$ and $g_o$ is given, Equation~\ref{posterior_dist_CCM_g} can be re-written as the following:

\begin{equation}  \label{posterior_dist_CCM_g2}
    P(g_c \vert \theta, g_{o})) \propto \left(\frac{1}{\vert c_{\phi}(\phi(g_c)) \vert} \right) P_{\phi}(\phi(g_c) \vert g_o, \theta).
\end{equation}

The conditional probability in Equation~\ref{posterior_dist_CCM_g2} is the probability distribution associated with a CCM, but restricted to $g_{o} \subset g_c$. Our approach to updating $g_u$ uses a MCMC Metropolis-Hastings (MH) algorithm nested within the Gibbs sampler to generate a series of networks, $g_1,\ldots,g_M$. As we are using an MCMC approach conditional on $\theta$, the normalizing constant $W_{\phi}(\theta)$ associated with $P_{\phi}(\phi(g_c) \vert g_o, \theta)$ is not needed. At each iteration $t$, the MCMC algorithm creates a proposal network, denoted as $gp_{t}$, by selecting an $E_{v_{i},v_{j}}$ to toggle, i.e., $E_{v_{i},v_{j}}$ is set to $0$ (removed) if it is currently in the network or set to $1$ (added) otherwise. At the end of the iteration, either the proposal is accepted ($g_t = gp_{t}$) or rejected ($g_t = g_{t-1}$) based on the following acceptance probability:

\begin{equation}
a_{ij} = \begin{cases}
0 & \mbox{ if } E_{v_i,v_j} = 1 \mbox{ for } g_o \hfill \\
r_{ij} & \mbox{ otherwise}, \\
\end{cases}
\end{equation}

\noindent where: 

\begin{equation}
r_{ij} = \min\left(1,\frac{P_{\mathscr{G}_n}(gp_{t} \vert \theta)}{P_{\mathscr{G}_n}(g_{t-1} \vert \theta)}\right).
\end{equation}

\noindent Approaches exist to estimate the ratio in $r_{ij}$.\citep{goyal2014sampling} In order to sample individuals $v_i$ and $v_j$ to select $E_{v_i,v_j}$, we use the tie-no-tie (TnT) method.\citep{morris2008specification} The TnT method samples two individuals with a link between them with probability $0.5$ and two individuals at random with probability $0.5$. The method is used to decrease the number of MCMC iterations until convergence.

\subsubsection{Update Parameters: $\theta$}\label{sec:update_CCM_parameters}

Based on the posterior distribution shown in Equation~\ref{eq:posterior_distr}, the full conditional distribution for $\theta$ is the following:\citep{koskinen2013bayesian} 

\begin{equation} \label{eq:update_theta}
    P(\theta \vert g_o \cup g_u) \propto \left(\frac{1}{\vert c_{\phi}(\phi(g_o \cup g_u)) \vert} \right) P_{\phi}(\phi(g_o \cup g_u) \vert \theta) \cdot \pi(\theta).
\end{equation}

\noindent As the size of the congruence class, $\vert c_{\phi}(\phi(g_o \cup g_u)) \vert$ does not depend on $\theta$, Equation~\ref{eq:update_theta} can be simplified to the following:

\begin{equation} \label{eq:update_theta_simple}
    P(\theta \vert g_o \cup g_u) \propto P_{\phi}(\phi(g_o \cup g_u) \vert \theta) \cdot \pi(\theta).
\end{equation}

\noindent Without the network normalizing term--the size of the congruence class--Equation~\ref{eq:update_theta_simple} can be evaluated using standard techniques for computing the product of two probability distributions. The simplification from Equation~\ref{eq:update_theta} to Equation~\ref{eq:update_theta_simple} provides our approach with the computational ability to investigate large networks. To further lessen computational complexity, we can select $P_{\phi}(\phi(g_o \cup g_u) \vert \theta)$ and $\pi(\theta)$ to be conjugate distributions. Sections~\ref{sec:simulation} and \ref{sec:application} provide examples of specifying these distributions.

As mentioned in Section~\ref{sec:background}, there are some practical considerations for specifying $P_{\phi}(\phi(g_o \cup g_u) \vert \theta)$. Due to the complexity of delineating $U_{\phi}$, it is more convenient for specifying distribution $Q_{\phi}(\theta)$ in Equation~\ref{eq:networkprob_static_pc_QW}. In contrast to updating the unobserved portion of the network ($g_u$) in the previous subsection, the value for $W_{\phi}(\theta)$ (as shown in Equation~\ref{eq:networkprob_static_pc_QW}) needs to be considered. 

Below we first provide a detailed description of an approach to estimate $W_{\phi}(\theta)$ using an MCMC procedure. Then we provide simulation results as well as theoretical and conceptual rationale to provide support that $W_{\phi}(\theta)$ may not be necessary to estimate for valid inference in particular settings--such as those we consider here. We can approximate $W_{\phi}(\theta)$ based on the following:

\begin{equation}
W_{\phi}(\phi(g) \vert \theta) = \vert U_{\phi}\vert * \frac{1}{M}\sum_{u \in U_{\phi,M}} Q_{\phi}(u \vert \theta),  
\end{equation}

\noindent where $\vert U_{\phi} \vert$ is the number of congruence classes in $U_{\phi}$, and $U_{\phi,M}$ is a random sample of $M$ elements from $U_{\phi}$. Recent advances in graph enumeration enable the estimation of $\vert U_{\phi} \vert$.\citep{goyal2022general} A random sample from $U_{\phi}$ can be obtained by assigning a uniform distribution for $P_{\phi}(\phi(g_o \cup g_u) \vert \theta)$, i.e., $P_{\phi}(\phi(g_o \cup g_u) \vert \theta) \propto 1$; illustrations of sampling uniformly from $U_{\phi}$ are provided in the Supplementary Materials.

It would not be necessary to estimate $W_{\phi}(\theta)$ if $W_{\phi}(\theta_1) \approx W_{\phi}(\theta_2)$ for $\theta$ in plausible ranges as the normalizing constant would cancel in the calculations. Finding theoretical bounds on $\frac{W_{\phi}(\theta_1)}{W_{\phi}(\theta_2)}$ for all settings would be quite difficult. Therefore, we conducted two simulations: one focused on degree distribution (property of interest in Section~\ref{sec:simulation}) and one focused on classification mixing matrix (property of interest for our investigation of Miami-Dade County, Section~\ref{sec:application}); see Supplement Section 1. Based on the simulation results, a 20\% change in $\theta$ only changed the estimated normalizing weight by at most 6.0\%.

Furthermore, for investigation of classification mixing matrices, we provide additional theoretical justification on why $W_{\phi}(\theta_1) \approx W_{\phi}(\theta_2)$ in particular settings; see Supplement Section 2. Theorem 1 in the Supplement asserts that all matrices with non-negative integers for entry $\DD{MM}{k,l}{}$ that are below particular values are graphical, i.e., valid classification mixing matrices. As social and sexual networks are sparse, the values for each entry, $\DD{MM}{k,l}{}$, should be sufficiently small compared to maximum possible value. Therefore, Theorem 1 indicates that $W_{\phi}(\theta_1) \approx 1$ is appropriate for some specifications of $P_{\phi}(\phi(g) \vert \theta)$; we use one of these specifications for $P_{\phi}(\phi(g) \vert \theta)$ in our investigation of disease dynamics in Miami-Dade County. Therefore, we can approximate $W_{\phi}(\theta)$ as $1$ for our analysis.

Finally, a conceptual argument for excluding $W_{\eta}(\theta)$ is that the update to $\theta$ would potentially have the desired distribution of network statistics. That is, the the distribution for $\theta$ as shown in Equation~\ref{eq:update_theta_simple} would be the product of $Q_{\phi}(\phi(g_o \cup g_u) \vert \theta)$ and $\pi(\theta)$, both quantities are specified by the investigator. We demonstrate this in the illustrative example in Section~\ref{sec:background} as well as in \cite{goyal2014sampling}.

Given our results from the simulation study (Supplementary Materials) investigating $W_{\phi}(\theta)$ as well as theoretical and conceptual arguments above, in the sections (Sections~\ref{sec:simulation} and~\ref{sec:application}) we do not include $W_{\phi}(\theta)$ in our updating of $\theta$. Even with this approximation, our broad set of simulations (Section~\ref{sec:simulation}) demonstrate large reduction in error. However, as these simulations and theoretical justification provide support for $W_{\phi}(\theta_1) \approx W_{\phi}(\theta_2)$ only in particular settings, additional research is necessary to develop more general results. We refer to this issue in the Discussion (Section~\ref{sec:discussion}). 

\section{Simulation Study} \label{sec:simulation}

To investigate our approach's ability to estimate network properties in the presence of missing sequence data, we conducted a simulation study that varies the proportion of observed data. The study focuses on estimating degree distributions for an entire network ($g_c$) based on a partially observed network ($g_o$). In Section~\ref{sec:application}, we investigate characteristic mixing matrix for MDC. 

To conduct the study, we simulate a population of $n=1000$ individuals and generate a network based on a CCM, wherein the degree distribution follows the same negative binomial distribution as used in the illustration in Section~\ref{sec:background}. The generated network represents a VGL network. To generate a partially observed network, we simulate the collecting and reporting of HIV sequences. To do so, we sample a fraction of the individuals. This sampling fraction represents individuals who have a sequence that was both collected and reported. We generate the partially observed network by constructing the induced subnetwork based on the sampled individuals; that is, two individuals are linked if and only if both are sampled, and are linked in the complete simulated network. The simulations use sampling fractions values from the following set: $\{0.1, \ldots, 0.9\}$.

\subsection{Model Specification}

The posterior distribution shown in Equation~\ref{eq:posterior_distr} requires specifying $P_{\mathscr{G}_n}(g_u \cup g_o \vert \theta)$ and $\pi(\theta)$. We assume $P_{\mathscr{G}_n}(g_u \cup g_o \vert \theta)$ can be modeled using CCM where $\theta$ are parameters associated with the degree distribution of the network; the probability distribution is shown below:

\begin{equation} \label{eq:networkprob_studyA}
P_{\mathscr{G}_n}(g_u \cup g_o) \vert \theta) =\left(\frac{1}{\vert c_{\phi}(\phi(g_u \cup g_o)) \vert} \right) P_{\phi}(\phi(g_u \cup g_o) \vert \theta).
\end{equation}

\noindent CCMs offer flexibility in defining the probability distribution for $P_{\phi}(\phi(g_u \cup g_o) \vert \theta)$. Here, we are able to assume that $P_{\phi}(\phi(g_u \cup g_o) \vert \theta)$ conforms to a multinomial distribution with parameter $\theta$, taking advantage of the CCM's  ability to represent a broad spectrum of degree distributions. Opting for a more constrained distribution, like a negative binomial, could potentially enhance the performance of our approach in simulation studies if this assumption proved correct. However, since the true functional form for the degree distribution of the VGL network associated with MDC is unknown, we prefer a more flexible distribution.

We posit that $\pi(\theta)$ follows a Dirichlet distribution with a parameter vector $\alpha_0$. To establish a non-informative prior, we set $\alpha_0$ uniformly to $\frac{1}{10000}$ for all components. The assumption that the distributions for $P(\phi(g_c) \vert \theta)$ (multinomial) and $\pi(\theta)$  (Dirichlet) are conjugate, which reduces the computational burden of updating CCM parameters.

\subsection{Results}

Below we present results for our Bayesian approach to estimation of the vector of parameters for the multinomial degree distribution. For each sampling fraction from $10\%$ to $90\%$, we perform the estimation procedure 100 times; hence, a total of 900 simulations. In the next subsection, we assess MCMC convergence for each of the $1000$ parameters in the multinomial distribution (representing the proportion of individuals with degree $0$ to $999$) as well as across the $900$ simulations. Section~\ref{sec:ess} provides results for the effective sample size for the MCMC iterations. Section~\ref{subseq:sim_bias} provides a comparison of the estimates derived from the partially observed networks and estimates based on the proposed imputation approach using the Hellinger distance as our comparison metric. 

\subsubsection{Convergence Results}

Our MCMC convergence assessment follows a methodology similar to that described in \cite{goyal2023estimating}. Initially, we analyze trace plots to identify an approximate iteration where the Markov chain reaches a stable state. Subsequently, guided by the potential convergence iteration identified, we conduct a statistical evaluation of convergence utilizing Geweke's convergence diagnostic.\citep{geweke1992evaluating} 

\begin{figure}
\centering
\includegraphics[width=400pt]{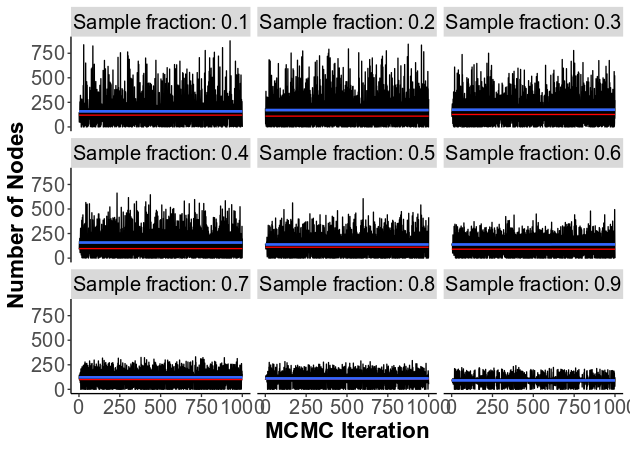}
\caption{Trace plots from simulations with varying values for sampling fraction $\{0.1, \ldots, 0.9\}$ are presented. Each plot illustrates the count of individuals having a degree of 2 (y-axis) across 1000 MCMC iterations (x-axis). Each trace plot displayed corresponds to the simulation with the median Mean Squared Error (MSE) value for the respective sampling fraction. Additionally, each plot includes a smooth (blue) curve across the points and a red line representing the value for the entire network.}
\label{fig:NetMissing_trace_degree}
\end{figure}

Figure~\ref{fig:NetMissing_trace_degree} presents trace plots from specific simulations, each employing varying sampling fractions ranging from ${0.1, \ldots, 0.9}$. These plots visualize the distribution of individuals with a degree of $2$ (y-axis) over $1000$ MCMC iterations (x-axis). Each trace plot corresponds to the simulation with the median Mean Squared Error (MSE) value for its respective sampling fraction. Additionally, each plot features a smooth (blue) curve across the points and a red line denoting the value for the entire network.

Importantly, none of the trace plots in Figure~\ref{fig:NetMissing_trace_degree} indicate any failure of the MCMC procedure to reach a stationary state. As anticipated, simulations with lower sampling fractions exhibit greater variance compared to those with higher sampling fractions. From a visual examination of a subset of trace plots, convergence is assessed to be attained around the $100^{th}$ iteration. 

We employ Geweke's z-score diagnostic to ascertain whether iterations $101-1000$ effectively represent samples from the stationary distribution. Specifically, we utilize this diagnostic to examine the equality of means between the initial 10\% of iterations and the latter 50\% of iterations of a Markov chain, discarding the initial $100$ iterations for burn-in as determined by the analysis of trace plots. This simulation study necessitates the evaluation of a total of $900,000$ chains, corresponding to $900$ simulations, with each estimating $1,000$ parameters for the multinomial distribution, one for each potential degree ranging from $0$ to $999$.

The majority of chains (94.7\%) exhibit either no or minimal variability, rendering the calculation of a Geweke z-score unfeasible. Among the chains where a z-score can be computed, the median absolute z-score is $0.9$, with the $25\%$ and $75\%$ quantiles at $0.4$ and $1.3$, respectively. Therefore, the majority of chains are determined to be in a stationary state for iterations $101-1000$. While according to Geweke's z-score, a small proportion of chains (<0.5\%) do not demonstrate a stationary state during iterations between $101-1000$, we expect that extending the runtime of these chains would enhance the performance of our approach. 

\subsubsection{Effective Sample Size Results} \label{sec:ess}

Restricted to the chains with variability, we have a median effective sample size of $901$  ($25\%$ and $75\%$ quantiles are $713$ and $1673$) using the CODA library in R.\citep{Rcoda, Rcran} Therefore, on average, we have samples that are effectively independent. We also see that many chains have a negative autocorrelation, indicating that they achieve a higher effective sample size compared to the number of MCMC iterations.

\subsubsection{Bias} \label{subseq:sim_bias}

\begin{figure}
\centering
\includegraphics[width=400pt]{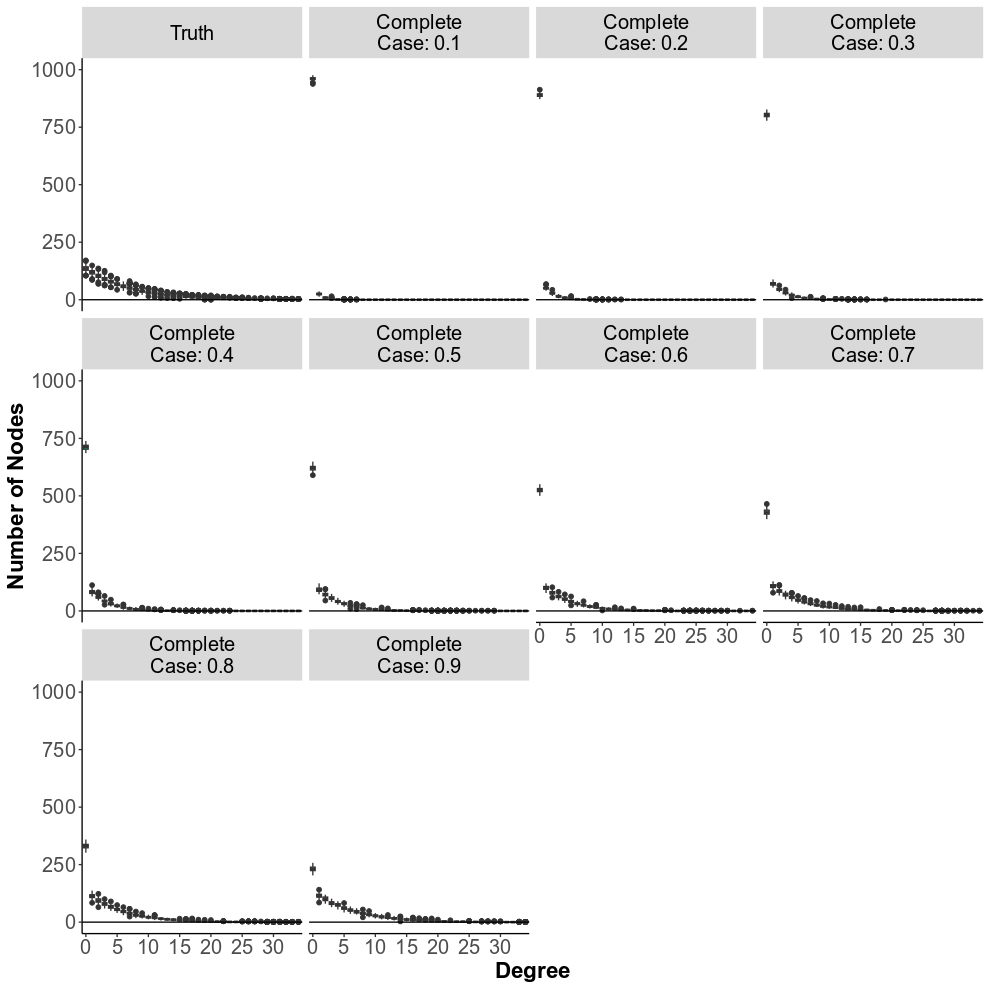}
\caption{Boxplots for the number of individuals with degrees 0 to 34. The first panel shows the number of individuals for the entire networks by degree. The remaining boxplots provide the number of individuals for the partially observed networks by degree for sampling fraction from $0.1$ to $0.9$ based on complete case analysis.}
\label{fig:NetMissing_boxplot_obs}
\end{figure}

Figure~\ref{fig:NetMissing_boxplot_obs} depicts boxplots for the number of individuals with degrees 0 to 34. The first panel shows the number of individuals by degree for the entire simulated network. The remaining boxplots provide the number of individuals for the partially observed network by degree for sampling fractions from $0.1$ to $0.9$ based on complete case analysis. As expected, the partially observed networks numbers have a larger number of individuals with smaller degrees compared to the entire network. For example, the proportion of individuals with degree $0$ for the entire network is $13.6\%$; however, when data are collected only from $50\%$ of the individuals, the proportion of individuals with degree $0$ is $62.0\%$.

\begin{figure}
\centering
\includegraphics[width=400pt]{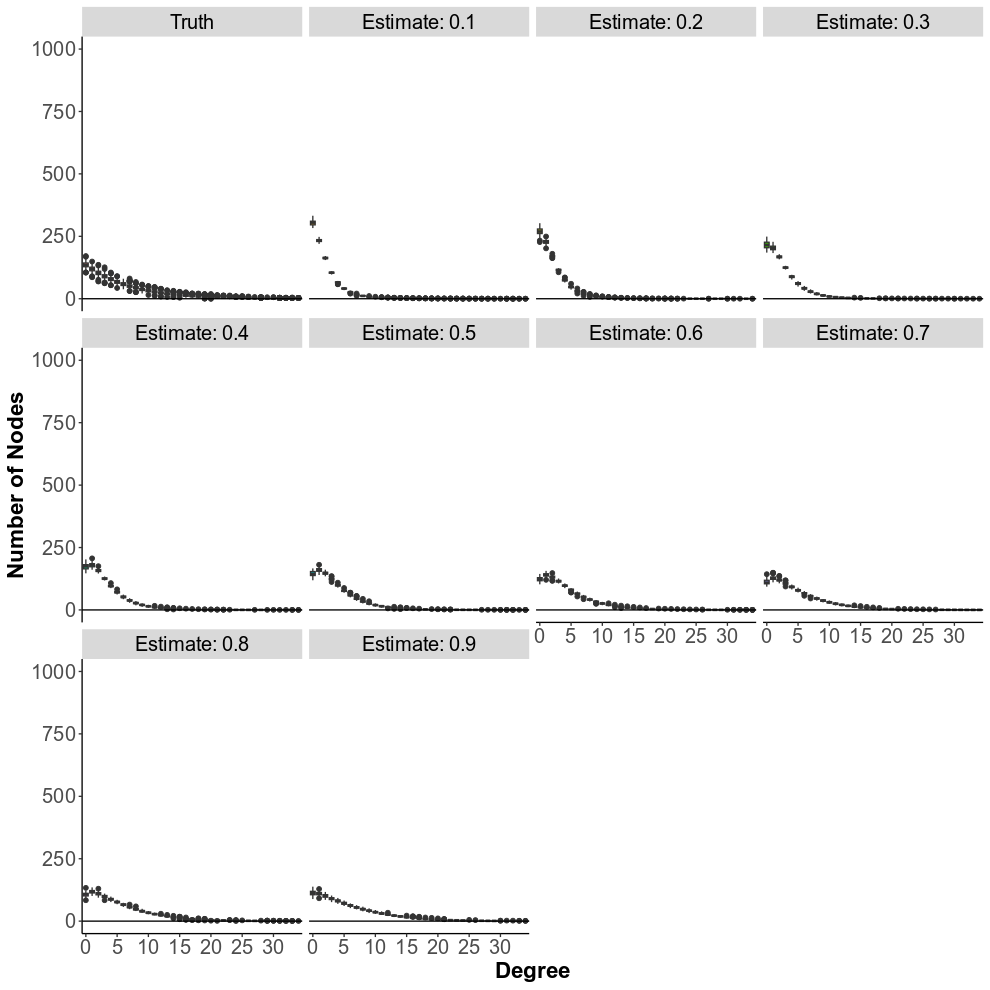}
\caption{Boxplots for the number of individuals with degrees 0 to 34. The first panel shows the number of individuals by degree for the entire network. The remaining boxplots provide our estimated number of individuals by degree for sampling fraction from $0.1$ to $0.9$.}
\label{fig:NetMissing_boxplot_estimate}
\end{figure}

Figure~\ref{fig:NetMissing_boxplot_estimate} provides information that is similar to that of Figure~\ref{fig:NetMissing_boxplot_obs}, but for estimated networks. The first panel shows the number of individuals for the entire networks by degree (identical to the first plot of Figure~\ref{fig:NetMissing_boxplot_obs}), whereas the remaining panels show the estimated number of individuals by degree for sampling fraction from $0.1$ to $0.9$. For example, our estimate for the proportion of individuals with degree $0$ is $14.5\%$ when data is collected from $50\%$ of the individuals, which is much closer to the entire simulated network ($13.6\%$) than are the values for the partially observed network ($62.0\%$).

In order to assess the improvement of our estimates compared to the observed networks, we calculate two sets of Hellinger distances--which provide a metric to assess the similarity between two probability distributions.\citep{huber1981robust} In our case, these distributions are degree distributions that describe the probability of a node have a particular degree. The first set of distances is between the  estimated degree distribution and the true degree distribution. The second set is between the observed degree distribution based on complete case analysis and the true degree distribution for the entire network. Table~\ref{table:bias_estimate_obs} shows the mean Hellinger distance for the observed and estimated networks for each sampling fraction. In addition, the table shows the mean reduction in this distance when using the estimated, compared to the observed, degree distributions. The reduction in error between our estimates and the quantities calculated based on complete case analysis from the partially observed network ranges from $43\%$ to $63\%$.

\begin{table}
\centering
\begin{tabular}{|c|c|c|c|}
  \hline
Sampling & Hellinger Distance & Hellinger Distance & Percent \\
Fraction & (Estimate) & (Complete Case) & Reduction\\
  \hline
  0.10 & 0.31 & 0.72 & 0.57 \\ 
  0.20 & 0.30 & 0.65 & 0.53 \\ 
  0.30 & 0.25 & 0.57 & 0.55 \\ 
  0.40 & 0.21 & 0.50 & 0.59 \\ 
  0.50 & 0.17 & 0.42 & 0.61 \\ 
  0.60 & 0.13 & 0.35 & 0.63 \\ 
  0.70 & 0.10 & 0.28 & 0.63 \\ 
  0.80 & 0.09 & 0.21 & 0.58 \\ 
  0.90 & 0.08 & 0.13 & 0.43 \\ 
   \hline
\end{tabular}
\caption{Mean Hellinger distance for the estimated networks and complete case analysis for each sampling fraction as well as the mean reduction in this distance using the estimated--compared to the observed--degree distribution.} \label{table:bias_estimate_obs}
\end{table}

\section{Miami-Dade County} \label{sec:application}

Miami-Dade County (MDC) has one of the highest rates of new HIV diagnoses in the nation; in 2022, the rate per 100,000 people in MDC was 39.3--nearly double that of Florida overall (20.6 per 100,000).\citep{FLHealthCharts} There are substantial disparities by race/ethnicity; non-Hispanic Blacks have a diagnosis rate of 69.3/100,000 compared to 17.8/100,000 for non-Hispanic Whites.\citep{FLHealthCharts} The epidemic is primarily concentrated within certain communities--notably among men who have sex with men (MSM), who accounted for 71\% of new HIV diagnoses in 2022 \citep{FLHealthCharts} but who comprise an estimated 6.6\% of the population of MDC.\citep{grey2016estimating}  In order to focus resources and assess the effectiveness of interventions on reducing transmissions, it is important to understand the disease dynamics in Miami-Dade County. 

Below, we investigate disease dynamics using MHS and focus exclusively on viral genetic networks among men who report having sex with men (MSM) in MDC because of their high disease burden. In particular, we investigate the nature of HIV spread across race/ethnicity and HIV transmission risk communities. Disparities in HIV risk can persist and intensify through preferential transmission among individuals within the same community.\citep{ragonnet2021sorting} Hence, it important to estimate the number of transmissions that are occurring within racial/ethnic categories or within transmission risk groups in order to identify the level of interventions necessary to mitigate the spread of HIV.   

\subsection{Data}

In order to investigate our two areas of research focus, we analyze data from MDC that includes demographic information on all PWH who resided in MDC at the end of 2021. We restrict our analysis to MSM. The database also contains the HIV sequences for this cohort. We define 4 categories of race/ethnicity (Hispanic, non-Hispanic Black , non-Hispanic White, and other) and 2 transmission risk categories among MSM: MSM with and without reported injection drug use (denoted as IDU and no IDU). 

Though it is standard clinical practice to collect viral sequences from people with HIV (PWH) at time of diagnoses, there is a fairly large proportion of PWH do not have a sequence.\citep{romero2020should} Based on these data, 30.9\% of PWH residing in MDC and identify as MSM at the end of 2021 have sequences that were collected and reported. The percentage varies by sociodemographic, transmission risk, and geography. See Table~\ref{tbl:MDC_demo} for characteristics of HIV-diagnosed MSM residing in MDC at the end of 2021 stratified by whether they have an HIV sequence that was collected and reported.

\begin{table} \label{tbl:MDC_demo}
\centering
\begin{tabular}{|c|c|c|c|}
  \hline
  **Characteristic** & No Sequence (N = 12,240) & Sequence (N = 5,463) & Percent Sequence\\
  \hline
Race &  &  &\\
Hispanic, any race & 7,661 / 12,240 (63\%) & 3,376 / 5,463 (62\%) & 30.6\% \\
Other & 163 / 12,240 (1.3\%) & 80 / 5,463 (1.5\%) & 32.9\% \\
Black & 2,294 / 12,240 (19\%) & 1,412 / 5,463 (26\%) & 38.1\% \\
White & 2,122 / 12,240 (17\%) & 595 / 5,463 (11\%) & 21.9\% \\
Transmission Risk &  &  &\\
MSM with no reported IDU& 11,749 / 12,240 (96\%) & 5,185 / 5,463 (95\%) & 30.6\% \\
MSM with reported IDU & 491 / 12,240 (4.0\%) & 278 / 5,463 (5.1\%) & 36.2\% \\
   \hline
\end{tabular}
\caption{Characteristics of HIV-diagnosed MSM residing in Miami-Dade County at the end of 2021 stratified by whether they have an HIV sequence collected and reported.}
\end{table}

\subsection{Construction of the VGL Network}

As in previous analyses, the VGL network is built by linking individuals with a genetic distance of 1.5\% or less;\citep{little2014using} this threshold has been found to identify direct or indirect transmissions.\citep{wertheim2017social} Pairwise distance was measured using the Tamura-Nei 93 algorithm. \citep{tamura1993estimation} All pairwise distances less than 1.5\% resulted in undirected edge between the pair of individuals with those sequences, regardless of when each member of the pair was sequenced. The missing sequences result in incomplete knowledge of the viral genetic network. 

\subsection{Transmissions Across Racial/ethnic and Transmission Risk Communities}

For our analysis, each individual has a race/ethnicity and transmission risk group. Based on both of these characteristics, each individual $i$ is assigned a category $m_i$ in $\{$Black, Hispanic, White, Other$\}$ $\times$ $\{$IDU, no IDU$\}$. Therefore, we have $q=8$ potential categories; for example, an individual can be assigned as a "Black MSM with not report injection drug use". We are interested in estimating the number of linkages among these $q$ categories; that is, estimating $\DDg{MM}{}{}{g}$, which is the $q \times q$ symmetric matrix representing the mixing by classification of the VGL network for MDC.

\subsection{Model Specifications}

For the posterior distribution shown in Equation~\ref{eq:posterior_distr}, we postulate a CCM for $P_{\mathscr{G}_n}(g_o \cup g_u) \vert \theta)$, where the network property of interest is $\DDg{MM}{}{}{g}$. For the PMF on the congruence classes, $P_{\phi}(\phi(g_o \cup g_u) \vert \theta)$, we specify the product of a Poisson distribution and a multinomial distribution. The Poisson distribution has parameter $\lambda$ that represents the mean number of linkages for the entire VGL network. The multinomial distribution has parameter $\theta$, where entry $\theta_{ij}$ represents the probability of a linkage between two individuals with characteristics $i$ and $j$. As indicated by Supplement Section 2, this specification of $P_{\mathscr{G}_n}(g_o \cup g_u) \vert \theta)$ will result in $W_{\phi}{\theta} \approx 1$. Therefore, we ignore $W_{\phi}(\theta)$ when updating $\theta$j within our Gibbs sampler. For our investigation, we specify the prior distributions $\pi_0(\lambda)$ and $\pi_0(\theta)$ as Gamma and Dirichlet distributions (conjugate priors) with parameters $\beta_0$ and $\alpha_0$, respectively. We set these parameters to result in a non-informative prior.

\subsection{Ignorability}

While it is not possible to know for certain the process for the collecting and reporting of HIV sequences in MDC, a plausible assumption is that this missing data mechanism is primarily based on individual-level covariates (e.g. race and transmission risk) and not on an individual's transmission patterns. Under this assumption, the missing data mechanism can be shown to meet the two criteria of ignorability discussed in Section~\ref{sec:ignorable}. Regarding the first, an unbiased estimate for the probability that  $I_{v_{i},v_{j}} = 1$ is the product of the probability of sampling $ v_{i}$ and $v_{j}$--both of these two probabilities can be estimated from the observed data. Therefore, the conditional distribution of $I_{v_{i},v_{j}}$ only depends on the observed part of data. For the second, the CCM characteristic matrix mixing parameters $\theta$ and the parameters for the missing data mechanism (i.e., sampling proportions based on observed covariates) are distinct, i.e., one set of parameters does not constrain the other. Therefore, the missing data mechanism is ignorable.  There is only one feature of the missing data mechanism of which authors are aware that may not be consistent with this assumption: sequences obtained through contact tracing. Even if contract tracing (i.e., multi-wave link-tracing) is an important aspect of the missing data mechanism, it has been shown to be ignorable of a missing data mechanism for ERGMs;\citep{handcock2010modeling} the same argument would hold for CCMs--as the parameters for the associated sampling process (proportion of nodes that are seed nodes and the number of waves) and the matrix mixing parameter, ($\theta$) are distinct.

\subsection{Results}

We first present results for the estimates of the classification mixing matrix as a whole and then present results by race/ethnicity and transmission risk groups. In the partially observed network, there are $5,696$ links among MSM compared to the estimate average number of $64,800$. Therefore, the partially observed network has only 8.8\% of the estimated number of links. Given that only 30.1\% of individuals have a sequence, we would expect to observe 9.6\% ($30.1\%^2$) of the edges. Therefore, our estimated total number of edges aligns with this expectation. We would not expect the numbers to be exactly the same given the imbalance of sequence coverage by race/ethnicity and transmission risk.

\begin{figure}
\centering
\includegraphics[width=400pt]{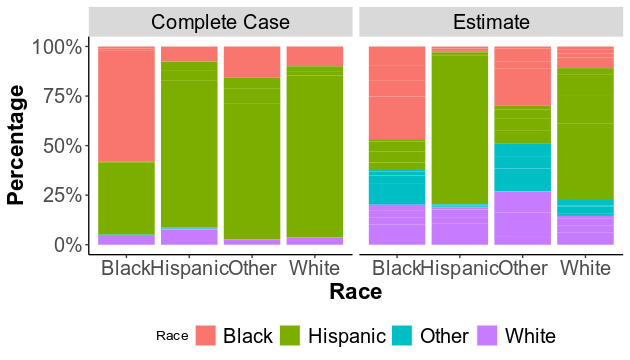}
\caption{Stack bars for the proportion of links between and among each race/ethnic group among MSM. The first panel is for the observed data. The second panel shows estimates using the approach presented.}
\label{fig:miami_mixing}
\end{figure}

To investigate HIV dynamics across racial/ethnic communities, we estimate the proportion of linkages from and between racial groups among MSM. In the partially observed network, viral genetic linkages from sequences from Black MSM have $57.9\%, 37.0\%, 4.6\%,$ and $0.6\%$ probability of connecting to a sequence from Black, Hispanic, White, or other race MSM, respectively. The corresponding estimates from our posterior distribution are $46.8\%, 15.4\%, 20\%,$ and $17.6\%$. Therefore, we estimate a lower than observed amount of linkage from Black to Black and Black to Hispanic and greater amount from Black to White and to other race. These shifts are potentially driven by the fact that Black MSM have the highest sequence coverage percentage (38.1\%) and White MSM have the lowest (21.9\%). We also observe large differences between the partially observed network and our estimates probabilities across the other races. Figure~\ref{fig:miami_mixing} depicts stack bars for the proportion of links between and among each race. The first panel refers to the partially observed network; and the second, shows estimates using our approach.

Our estimates of mixing differ than what would be expected under random mixing across racial/ethnic communities. Under random mixing, we would expect the proportion of linkages to be equal to the proportion of individuals within the racial/ethnic community. Therefore, we would expect $20.9\%$, $62.3\%$, $1.4\%$, and $15.3\%$ of edges to link to a Black, Hispanic, other, and White PLW, respectively, as these are the proportion of individuals in the respective communities (see Table~\ref{tbl:MDC_demo}). 

\begin{figure}
\centering
\includegraphics[width=400pt]{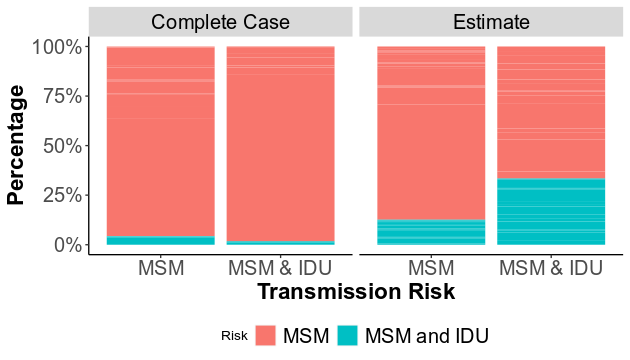}
\caption{Stack bars for the proportion of links within and between each transmission risk category (MSM with no IDU, MSM with reported IDU). The first panel shows the observed data; the second, estimates using our approach.}
\label{fig:miami_mixing_risk}
\end{figure}

As with race/ethnicity, we estimate the proportion of linkages within and across transmission risk groups among MSM, see Figure~\ref{fig:miami_mixing_risk}. The first panel shows the observed data; the second, shows estimates using our approach. Comparing the two panels, we see an increase in the number of linkages associated with individuals that are MSM with reported IDU. In particular, the incerease in linkages from MSM with no reported IDU compared to MSM with reported IDU ranges from $4.4\%$ to $12.2\%$. Increase in linkages from MSM with reported IDU range from $1.7\%$ to $32.3\%$. Under random mixing by transmission risk, we would expect $4.3\%$ of edges to link to MSM with reported IDU (see Table~\ref{tbl:MDC_demo}). 

\section{Discussion} \label{sec:discussion}

Our approach allows Bayesian inference of properties of large networks based on partially observed networks. For setting with large network size, prior to the development of our approach, only complete case analysis--which assume that the entire network of interest was observed–could be undertaken. In the paper, we investigate HIV transmission patterns among MSM in Miami-Dade County within and between racial/ethnic and injection drug use risk categories. Our results indicate large differences between accounting for missing data compared to conducting complete-case analyses. In particular, we find higher levels of mixing among MSM across racial/ethnic categories. Our findings are similar for transmission risk (defined by reported injection drug use) groups. Furthermore, our estimates of mixing differ from what would be expected under the assumption that viral genetic linkage occurred at random by race/ethnicity or transmission risk. Specifically, both the analyses accounting for missing data and complete-case analyses show evidence of assortativity in mixing by race/ethnicity, with a higher proportion of within group linkages observed compared to if individuals selected partners at random. However, the analysis accounting for missing data slightly attenuated this finding, observing more linkages between racial/ethnic groups compared to the complete-case analysis. For example, the complete case analysis indicated few linkages between Black MSM and White MSM (<5\%), whereas this increased to 20\% in our analysis accounting for missing data. We observed a similar increase in linkages between Hispanic MSM and White MSM, highlighting the potential importance of these mixing patterns on HIV transmission. 

Our findings have potential implications for designing and implementing HIV prevention programs as well as how to evaluate their effectiveness. For example, preferential mixing can be manifest as the presence of ``communities'', i.e., groups of entities that are highly-interconnected with each other and loosely connected with other entities outside their group. The cohesiveness of communities affects the spread of disease,\citep{huang2007epidemic} as well as which strategies would be effective on mitigating the spread.\citep{salathe2010dynamics} Our analysis indicates that there are likely more transmissions occurring between MSM of different race/ethnic groups than indicated with previous molecular HIV surveillance analysis not accounting for missing data. As such, for example, interventions to reduce HIV among Black or Hispanic MSM should additionally focus on other groups such as White MSM or MSM of other races/ethnicities. Conversely, interventions among White MSM could have a greater impact than previously estimated on transmission among Black or Hispanic MSM. 

The strength of our work is that our proposed methods allow us to conduct Bayesian inference for complex network properties for large networks with missing data; nonetheless, there is a need to further develop statistical methods in several areas. The first is estimating bounds for $W_{\phi}(\theta_1) / W_{\phi}(\theta_2)$ for $\theta_1$ and $\theta_2$ in plausible values. A second limitations is assessing convergence of MCMC algorithms when modeling a large number of network model parameters,\citep{rajaratnam2015mcmc} which is possible for CCMs and demonstrated in our simulation studies and investigating disease dynamics for Miami-Dade County. Another area of further research is developing methods that allow further inference on incomplete network data, including statistical properties of the face-value likelihood for networks. A fourth area is develop of methods to determine the correct network model--both network properties to include and their functional form--in the presence of missing data. 

In conclusion, we identify important disease dynamics of the HIV epidemic among MSM in Miami-Dade County using a novel approach to address missing network data. Using the Bayesian paradigm, our approach overcomes the high computational burden of previous model-based approaches for estimating network properties from partially observed networks. This capability allowed us to make formal statistical inferences regarding complex network properties of the VGL network associated with Miami-Dade County that were previously not feasible. Overall, our work provides an important methodological foundation for drawing inference from molecular epidemiological and other network data to inform HIV prevention and beyond.

\section*{Acknowledgments}

This research is supported by grants from the National Institutes of Health (R01 AI-147441, R01 MH-132151 and P30 AI-036214). Conflict of Interest: SJL has received funding from Gilead Sciences paid to her institution. NKM receives unrestricted research grants from Gilead and AbbVie unrelated to this work.

\bibliographystyle{imsart-nameyear}  
\bibliography{references}  

\end{document}